\documentclass[english,american]{article}
\usepackage[T1]{fontenc}
\usepackage[utf8]{inputenc}
\usepackage{parskip}
\usepackage{amsmath}
\usepackage{amssymb}
\usepackage{setspace}
\makeatletter

\providecommand{\tabularnewline}{\\}

\usepackage{babel}

\makeatother

\usepackage{babel}
\begin{document}
\title{Integrating Network Psychometrics and LLMs: The Ising-Embeddings-Model
applied to Reliability Auditing}
\author{Matthias von Davier\thanks{The author acknowledges the use of generative AI tools for assistance
with code development (vibe-coding) and light editing of the manuscript;
the code produced by the LLM was checked by the author of the submission.
All conceptual ideas, mathematical derivations, were developed and
the empirical analyses were developed by the author of the submission.}}
\maketitle
\begin{abstract}
Scoring consistency for constructed-response items in large-scale
assessments is typically estimated through double-scoring, which uses
small samples and assumes independence among responses. We present
an integrated framework combining network psychometrics with the Linguistic-Integrated
Reliability Audit (LiRA) via a modified Ising model. The model defines
a joint distribution over binary correctness labels with pairwise
interactions set to the cosine similarity of sentence embeddings and
a global bias parameter for item difficulty. LiRA's weighted majority
voting over semantic neighborhoods is shown to approximate the conditional
logistic distributions of this Ising model. The parametric framework
supports benchmark score generation, uncertainty quantification, and
missing label imputation; parameters are estimated by maximum pseudo-likelihood.
The approach uses the full dataset without requiring extensive double-scoring,
accounts for semantic dependencies, and provides diagnostics for rater
inconsistencies. The integration of LiRA's scalable methodology with
a probabilistic graphical model offers a comprehensive tool for reliability
assessment in international assessments such as PIRLS, PISA, and TIMSS. 

\end{abstract}

\section{Introduction}

This work develops a model-based justification and extension of the
method for creating a reference score in the LiRA framework. In LiRA,
a small number of responses are selected that show maximum similarity
to the target response in order to create a reference score that is
used to evaluate the agreement between the score assigned to the target
response and that reference score.

This approach weights the reference response by the similarity between
the target and comparison responses. Here we show that this approach
can be justified by means of the Ising model, which was introduced
to psychometrics by the prolific Amsterdam school.

\subsection{Network Psychometrics and the Ising Model}

Network psychometrics models the interdependence among observed variables
without reliance on latent variables. A typical setup consists of
$K$ observed variables $X_{k}\in\left\{ 0,1\right\} $ with $k=1,\dots,K$,
and the assumption that the joint distribution of $\left(X_{1},\dots,X_{K}\right)$
exists and can be modeled using a parametric probability distribution
$P\left(X_{1},\dots,X_{K}\right)$ for $K$-dimensional binary variables.
Note that instead of Boolean $x\in\left\{ 0,1\right\} $ variables,
the Ising model often utilizes a notation 
\[
X^{*}_{k}=2X_{k}-1\in\left\{ -1,1\right\} 
\]
that reflects its origin in statistical mechanics, as it was originally
developed to describe magnetic spins (Ising, 1925). We will use $X_{k}$
and $X^{*}_{k}$ interchangeably in the text but will use the appropriate
parametrization in the equations as needed.

In Ising's approach, the probability of a realization (a state) $\left(x_{1},\dots,x_{K}\right)$
of the $X_{k}$ is modeled as 
\[
P\left(x_{1},\dots,x_{K}\right)=\frac{\exp\left[-\beta H\left(x_{1},\dots,x_{K}\right)\right]}{Z}
\]
where 
\[
Z_{\beta}=\sum_{\left(y_{1},\dots,y_{k}\right)\in\left\{ 0,1\right\} ^{K}}\exp\left[-\beta H\left(y_{1},\dots,y_{k}\right)\right]
\]
is the partition function, a sum over all possible states. Finally,
$H$ is a function referred to as the Hamiltonian and is defined as
\[
H\left(x_{1},\dots,x_{K}\right)=-\frac{1}{2}\sum_{j\ne l}w^{*}_{jl}x^{*}_{j}x^{*}_{l}-\mu\sum_{k}v_{k}x^{*}_{k}.
\]
The weights $w_{jl}$ form a symmetric $K\times K$ matrix, with $w_{jl}=w_{lj}$,
and the factor $\frac{1}{2}$ is necessary to avoid double counting
when summing over all pairs $j\ne l$. The parameters $\beta$ and
$\mu$ are referred to as the temperature and magnetic moment, respectively.
It should be noted that the expression for $P_{\beta}$ can be rewritten
as 
\[
P_{\beta}\left(x_{1},\dots,x_{K}\right)=\frac{\exp\left[\beta\sum_{j\ne l}w^{*}_{jl}x^{*}_{j}x^{*}_{l}+\beta\mu\sum_{k}v_{k}x^{*}_{k}\right]}{\sum_{\left(y_{1},\dots,y_{k}\right)\in\left\{ 0,1\right\} ^{K}}\exp\left[\beta\sum_{j\ne l}w^{*}_{jl}x^{*}_{j}x^{*}_{l}+\beta\mu\sum_{k}v_{k}y^{*}_{k}\right]}
\]
by defining $w_{jl}=\frac{1}{2}w^{*}_{jl}$ and inserting the definition
above into the expression for $P\left(x_{1},\dots,x_{K}\right)$.

The Ising model's moment of fame in psychometrics came when it was
claimed that this probability distribution for $K$ binary variables
is very similar to that of the conditional Rasch model (Molenaar,
2004). Instead of assuming a latent variable $\theta$ and conditional
independence of the $X_{k}$ given $\theta,$ the promise of the Ising
model was to provide a probabilistic description of a number of binary
variables that relies only on weighted pairwise interactions $w_{jl}x^{*}_{j}x^{*}_{l}$
and the (weighted) sum of responses $\sum_{k}\nu_{k}x^{*}_{k}$. Research
eagerly picked up this approach as an alternative to classical IRT
models (Epskamp et al., 2018).

However, replacing the parsimonious and arguably plausible assumptions
of the Rasch model (von Davier, 2016) with those of the Ising model
comes at a cost: The partition function $Z$ grows rapidly with growing
$K$, as it involves $2^{K}$ function evaluations. Another challenge
is that the fully estimated Ising model requires $K\left(K-1\right)/2$
interaction parameters $w_{jl}$, plus $K$ bias parameters $v_{k}$,
plus two parameters $\beta,\mu$, so that the parameter count essentially
grows quadratically with $K$, while the Rasch model requires parameters
that grow linearly with $K$.

von Davier (2018) suggests using noise contrastive estimation (Gutmann
\& Hyvärinen, 2012) to get around the estimation issue involving $Z$,
but estimation of the model parameters of the unconstrained Ising
model remains a challenge. In this paper, we will adopt a pseudo-likelihood
approach for a reduced model that is expected to provide accurate
and fast results even for large $K.$

\subsection{Linguistic-Integrated Reliability Audit (LiRA)}

In LiRA, we assume that there is a response variable $Y_{i}$ that
encodes the answers provided by a sample of $K$ test-takers to the
$i$-th question on a test. Assume these are text-based (written)
responses that need to be classified as either correct $(=1)$ or
incorrect $(=0)$, either by human raters, by a machine learning algorithm,
or, more recently, by an LLM (Jung et al., 2026).

Scoring can be viewed as a function that maps the $K$ written responses
\[
f_{M}:\left(y_{1},\dots,y_{K}\right)\mapsto\left(X_{M1},\dots,X_{MK}\right)\in\left\{ 0,1\right\} ^{K}
\]
where $f_{M}$ stands for a specific approach ($M=H_{a}$ may denote
a human rater, $M=C_{b}$ an ML algorithm used for scoring, and $M=A_{c}$
an AI {[}LLM{]} scorer). Scoring agreement assessment involves using
two different mappings and calculating a divergence measure, e.g.,
\[
0\le EA\left[f_{H_{a}}\left(y_{1},\dots,y_{K}\right),f_{C_{b}}\left(y_{1},\dots,y_{K}\right)\right]=\frac{\#\left\{ f^{[k]}_{H_{a}}\left(y_{k}\right)=f^{[k]}_{C_{b}}\left(y_{k}\right)\right\} }{K}\le1
\]
where $X_{Mk}=f^{[k]}_{M}\left(\cdot\right)$ refers to the $k$-th
component of the image vector. The first expression gives the proportion
of exact agreement, while 
\[
0\le\kappa\left(f_{H_{a}}\left(y_{1},\dots,y_{K}\right),f_{C_{b}}\left(y_{1},\dots,y_{K}\right)\right)\le1
\]
gives Cohen's kappa (Cohen, 1960).

Linguistic similarity can be assessed by mapping the written responses
$y_{k}$ using sentence embeddings (e.g. Feng et al., 2022) or even
a Bag-of-Words (BoW) approach. The current approach used in LiRA is
based on multilingual embeddings (Bezirhan, Jung \& von Davier, 2026).
Under this approach, each raw response is transformed into a latent
variable representation 
\[
emb\left(y_{k}\right)=e_{k}\in\mathbb{R}^{d}
\]
of length $d$, which represents the dimension of the embedding. Typical
dimensions are $d>500.$ For example, Microsoft's mE5-large maps sentences
into 1024 dimensions, while Google's LaBSE maps sentences into 768
dimensions (Wang et al., 2024; Feng et al., 2022). After all responses
$y_{1},\dots,y_{K}$ are mapped into the latent variable space, the
embeddings $\left(e_{1},\dots,e_{K}\right)$ are assessed for similarity,
independent of how they were scored. A popular approach is to use
the cosine similarity (Gomaa \& Fahmy, 2013), which gives 
\[
s_{kl}=sim\left(e_{k},e_{l}\right)=\frac{e_{k}e_{l}}{\left\Vert e_{k}\right\Vert \left\Vert e_{l}\right\Vert }=sim\left(e_{l},e_{k}\right)\in\left]-1,1\right[.
\]
The cosine similarities of all pairs of embeddings form a symmetric
$K\times K$ matrix 
\[
Sim=\left[\begin{array}{cccc}
1 & \frac{e_{1}e_{2}}{\left\Vert e_{1}\right\Vert \left\Vert e_{2}\right\Vert } & ... & \frac{e_{1}e_{K}}{\left\Vert e_{1}\right\Vert \left\Vert e_{K}\right\Vert }\\
\frac{e_{2}e_{1}}{\left\Vert e_{2}\right\Vert \left\Vert e_{1}\right\Vert } & 1 & ... & \frac{e_{2}e_{K}}{\left\Vert e_{2}\right\Vert \left\Vert e_{K}\right\Vert }\\
... & ... & 1 & ...\\
\frac{e_{K}e_{1}}{\left\Vert e_{K}\right\Vert \left\Vert e_{1}\right\Vert } & \frac{e_{K}e_{2}}{\left\Vert e_{K}\right\Vert \left\Vert e_{2}\right\Vert } & ... & 1
\end{array}\right]
\]
which contains all similarities twice, mirrored around the diagonal.

The LiRA approach as proposed by Jung et al. (2026) uses this matrix
$Sim$ as follows: 
\begin{enumerate}
\item Starting with $k=1$: For each response $y_{k}$ and associated embedding
$e_{k}$, find the $h$ most similar embeddings among the remaining
responses, using the row $Sim\left[k\right]\backslash k$, i.e., excluding
the diagonal element, which trivially equals 1. 
\item Record the similarities for the $h$ most similar responses, multiply
each by the score assigned to the corresponding response, and divide
the sum of these products by the sum of the recorded similarities.
This provides a weighted reference score. 
\item Compare this reference score based on the weighted scores of the most
similar responses to the score assigned to $y_{k}$. 
\item Repeat for $k+1$. 
\end{enumerate}
This algorithm creates $K$ reference scores $\hat{r}_{k}=g\left(x_{r_{1}(k)},x_{r_{2}(k)},\dots,x_{r_{h}(k)}\right)$
using the scores assigned to the $h$ most similar responses $x_{r_{j}(k)}$,
excluding the target response itself. This imputed reference score
$\hat{r}_{k}$ can be based on the same scoring engine and provides
a measure of scoring consistency similar to a scoring reliability
assessment with multiple scorers, or a reference score based on expert
judgment.

This approach can be understood as a non-parametric approach to imputation,
such as mean replacement or predictive mean matching (Rubin, 1987),
in that it uses simple statistics calculated from a sub-sample of
other observations and replaces the unknown reference scorer's response
with that statistic.

\section{Fusing Network Psychometrics and Linguistic Similarity-Based Scoring
Reliability}

The central idea in this article is to modify the Ising model so that
it remains a probability function for a set of $K$ binary variables
$\left(x_{1},\dots,x_{K}\right)$, but is applied to the human- or
AI-generated scores given to $K$ written responses $\left(y_{1},\dots,y_{K}\right)$,
for which we assume that 
\[
w_{jl}\sim sim\left(e_{j},e_{l}\right)
\]
where $w_{jl}=sim\left(e_{j},e_{l}\right)\in\left]-1,1\right[$ is
the corresponding weight, and $g\left[z\right]$ with $z=\left(x^{*}_{j},x^{*}_{l}\right)$
is an activation function. Examples are $g\left[z\right]=x^{*}_{j}x^{*}_{l},$
as well as $g\left[z\right]=\mathrm{LeakyReLU}\left[x^{*}_{j}x^{*}_{l},\alpha\right]$.
We also include a simplified bias function in the form of 
\[
\mu\sum_{k}h_{k}\left(x^{*}_{k}\right),
\]
where typically $h_{k}\left[x^{*}_{k}\right]=\mathrm{ReLU}\left(x^{*}_{k}\right)=x_{k}$.
This ensures that the parameter $\mu$ is related to how many responses
were scored as correct in the sample.

The rationale for using activations such as $\mathrm{LeakyReLU}\left[z\right]$
is somewhat different from why they are used in multilayer neural
networks. For example, 
\[
\mathrm{lR}_{\alpha}\left(x^{*}_{j}x^{*}_{l}\right)=\mathrm{LeakyReLU}\left[x^{*}_{j}x^{*}_{l},\alpha\right]=\begin{cases}
1 & x_{j}=x_{l}\\
-\alpha & \text{otherwise}
\end{cases}
\]

Here, this activation function assigns different weights depending
on whether scores agree or disagree. Disagreement can be weighted
higher, $\alpha>1,$ or lower than agreement, $\alpha<1$. However,
by default one can assume that agreement and disagreement are equally
weighted, i.e., $\alpha=1$.

\subsection{Integrating the Ising Model and LiRA}

In the Ising model, the argument of the probability function is given
by 
\[
I\left(x^{*}_{1}\dots,x^{*}_{K},\mathbf{w},\mathbf{v},\beta,\mu\right)=\beta\sum_{j\ne l}\frac{1}{2}w_{jl}x^{*}_{j}x^{*}_{l}+\beta\mu\sum_{k}v_{k}x^{*}_{k}
\]

where $\beta,\mu>0$. Assuming non-negative weights, this function
of the state $\left(x^{*}_{1},\dots,x^{*}_{K}\right)\in\left\{ -1,1\right\} ^{K}$
is maximized for a given set of weights $\left(w_{jl}\right)_{j\ne l:1,...K}$
whenever pairs with the largest weights $w_{jl}>0$ agree and pairs
with small weights disagree. It appears that, empirically, pairs of
sentence embeddings are rarely negatively related when calculating
their cosine similarity. This may be a limitation of how the embedding
space spans the real numbers, but if this is indeed the case, this
feature can be used in the LiRA Ising model.

With the above approximate equating of the weights $w_{jl}=sim\left(e_{j},e_{l}\right)$
to the similarity between responses, we generate a weight matrix that
has large positive entries for highly similar responses, which should,
on average, receive the same score. For dissimilar responses, this
means their cosine similarity is still positive, but closer to zero,
whereas highly similar responses, by design of sentence embeddings,
should yield similarity measures close to 1. This approach assumes
that dissimilar responses are more likely to be scored differently,
while more similar responses are assumed to be scored the same (both
correct or both incorrect). For the term related to the bias, we can
simplify it by setting $\lambda=\beta\mu$ to avoid having a multiplicative
expression.

Taken together, the simplified Ising model can be written as 
\[
P\left(x_{1},\dots,x_{K}\right)=\frac{\exp\left[\beta\sum_{j\ne l}\frac{1}{2}sim\left(e_{j},e_{l}\right)x^{*}_{j}x^{*}_{l}+\lambda\sum_{k}x_{k}\right]}{Z}
\]
to express the probability of a set of scores $\left(x_{1},\dots,x_{K}\right)$
produced by humans, an algorithm, or a language model. The main difference
from the fully parameterized Ising model is that we do not estimate
the $w_{jl}$. Instead, we assume that the similarity measures $e_{k}$
can be used to model these weights via $w_{jl}=sim\left(e_{j},e_{l}\right)$.
Under this assumption, the probability of a set of $K$ scores can
be expressed as above, with $\beta$ and $\mu$ as the model parameters.

\subsection{Relationship to Logistic Regression}

For any $k\in\left\{ 1,\dots,K\right\} $, the probability of $x_{k}$
given the other scoring variables $R_{k}=\left\{ x_{j}:j\ne k\right\} $
equals 
\[
P\left(x_{k}=1|x_{j}:j\ne k\right)=\frac{P\left(x_{1},\dots,x_{k}=1,\dots,x_{K}\right)}{P\left(x_{1},\dots,x_{k}=0,\dots,x_{K}\right)+P\left(x_{1},\dots,x_{k}=1,\dots,x_{K}\right)}
\]
and further 
\[
P\left(x_{k}=1|x_{j}:j\ne k\right)=\frac{1}{1+\frac{P\left(x_{1},\dots,x_{k}=0,\dots,x_{K}\right)}{P\left(x_{1},\dots,x_{k}=1,\dots,x_{K}\right)}}.
\]

For the ratio in this expression, we have 
\[
\frac{P\left(x_{1},\dots,x_{k}=0,\dots,x_{K}\right)}{P\left(x_{1},\dots,x_{k}=1,\dots,x_{K}\right)}=\frac{\exp\left[\beta\sum^{K}_{j=1;j\ne k}-\frac{1}{2}sim\left(e_{j},e_{k}\right)x^{*}_{j}+\mu\left[\sum^{K}_{j=1;j\ne k}x_{j}\right]\right]}{\exp\left[\beta\sum^{K}_{j=1;j\ne k}\frac{1}{2}sim\left(e_{j},e_{k}\right)x^{*}_{j}+\mu\left[1+\sum^{K}_{j=1;j\ne k}x_{j}\right]\right]}
\]
since the partition function cancels in the ratio, as do all terms
that do not involve $k$, because these are identical in the numerator
and denominator. The summation is defined as $\sum^{K}_{j=1;j\ne k}x_{j}=\sum^{K}_{j=1}x_{j}0_{\left\{ k\right\} }\left(j\right)$
with $0_{\left\{ y\right\} }\left(x\right)=1-1_{\left\{ y\right\} }\left(x\right)$
and contains all summands except the $k-th.$ Further, we find 
\[
\frac{P\left(x_{1},\dots,x_{k}=0,\dots,x_{K}\right)}{P\left(x_{1},\dots,x_{k}=1,\dots,x_{K}\right)}=\exp\left[\beta\sum^{K}_{j=1;j\ne k}-sim\left(e_{j},e_{k}\right)x^{*}_{j}-\mu\right].
\]
Finally, we obtain the probability function 
\[
P\left(x_{k}=1|x_{j}:j\ne k\right)=\frac{1}{1+\exp\left[\beta\left(\sum^{K}_{j=1;j\ne k}-sim\left(e_{j},e_{k}\right)x^{*}_{j}\right)-\mu\right]}
\]
which equals 
\[
P\left(x_{k}=1|x_{j}:j\ne k\right)=\frac{\exp\left[\beta\left(\sum^{K}_{j=1;j\ne k}sim\left(e_{j},e_{k}\right)x^{*}_{j}\right)+\mu\right]}{1+\exp\left[\beta\left(\sum^{K}_{j=1;j\ne k}sim\left(e_{j},e_{k}\right)x^{*}_{j}\right)+\mu\right]}
\]
for each of the response score variables $x_{k}$ when assuming the
Ising model for the full set of scored responses $\left(x_{1},\dots,x_{K}\right)$.

This is a very useful result because it states that we can predict
the scores $x_{k}$ that the individual responses $y_{k}$ received
based on all other response scores $\left\{ x_{j}:j\ne k\right\} $
and the similarities between those responses and the target response
$sim\left(e_{j},e_{k}\right)$ when the response process is governed
by the Ising model with weights equal to scaled similarities $sim\left(e_{j},e_{k}\right)$.

\subsection{Estimation}

The similarities $sim\left(e_{j},e_{k}\right)$ will be taken as known
constants, calculated from a (multilingual) embedding model such as
those described by Bezirhan et al. (2026) and Wang et al. (2024).
Thus, the only parameters to be determined are the temperature $\beta$
and the bias $\mu$ in the modified Ising model. A straightforward
way to perform estimation of these two parameters, that avoids calculation
of the partition function $Z,$ which is practically intractable for
large $K$, is outlined in this section.

For a more compact notation, we use 
\[
C_{k}=\left(\sum^{K}_{j=1;j\ne k}sim\left(e_{j},e_{k}\right)x^{*}_{j}\right)=-\left(-\sum^{K}_{j=1;j\ne k}sim\left(e_{j},e_{k}\right)x^{*}_{j}\right).
\]
Then we have 
\[
P\left(x_{k}=1|x_{j}:j\ne k\right)=\frac{\exp\left[\beta C_{k}+\mu\right]}{1+\exp\left[\beta C_{k}+\mu\right]}=\frac{1}{1+\exp\left[-\beta C_{k}-\mu\right]}
\]
and 
\[
P\left(x_{k}=0|x_{j}:j\ne k\right)=\frac{1}{1+\exp\left[\beta C_{k}+\mu\right]}=\frac{\exp\left[-\beta C_{k}-\mu\right]}{1+\exp\left[-\beta C_{k}-\mu\right]}.
\]

Pseudo-likelihood estimation is an alternative that can be applied
here because we have independent scores $x_{k}$, and we use embeddings
from a pretrained model to calculate $w_{jk}=\frac{1}{2}sim\left(e_{j},e_{k}\right)$.
We obtain the pseudo-likelihood function 
\[
L_{p}=\sum^{K}_{k=1}\ln\left[\left(\frac{1}{1+\exp\left[-\beta C_{k}-\mu\right]}\right)^{x_{k}}\left(\frac{1}{1+\exp\left[\beta C_{k}+\mu\right]}\right)^{1-x_{k}}\right]
\]
and further 
\[
L_{p}=-\sum^{K}_{k=1}x_{k}\ln\left(1+\exp\left[-\beta C_{k}-\mu\right]\right)+\left(1-x_{k}\right)\ln\left(1+\exp\left[\beta C_{k}+\mu\right]\right).
\]

The derivative of the pseudo-likelihood with respect to $\beta$ is
found as 
\[
\frac{\partial}{\partial\beta}\ln\left(1+\exp\left[-\beta C_{k}-\mu\right]\right)=-C_{k}\frac{\exp\left[-\beta C_{k}-\mu\right]}{1+\exp\left[-\beta C_{k}-\mu\right]}=-C_{k}\frac{1}{1+\exp\left[\beta C_{k}+\mu\right]}
\]
and 
\[
\frac{\partial}{\partial\beta}\ln\left(1+\exp\left[\beta C_{k}+\mu\right]\right)=C_{k}\frac{\exp\left[\beta C_{k}+\mu\right]}{1+\exp\left[\beta C_{k}+\mu\right]}.
\]
Therefore 
\[
\frac{\partial L_{p}}{\partial\beta}=\sum^{K}_{k=1}C_{k}\left[x_{k}\frac{1}{1+\exp\left[\beta C_{k}+\mu\right]}-\left(1-x_{k}\right)\frac{\exp\left[\beta C_{k}+\mu\right]}{1+\exp\left[\beta C_{k}+\mu\right]}\right]
\]
or alternatively 
\[
\frac{\partial L_{p}}{\partial\beta}=\sum^{K}_{k=1}C_{k}\left[x_{k}P\left(x_{k}=0|x_{j}:j\ne k\right)-\left(1-x_{k}\right)P\left(x_{k}=1|x_{j}:j\ne k\right)\right].
\]

Intuitively, this gradient approaches zero if we obtain a $\beta$
that minimizes $P\left(x_{k}=0|x_{j}:j\ne k\right)$ whenever $x_{k}=1$
and minimizes $P\left(x_{k}=1|x_{j}:j\ne k\right)$ whenever $x_{k}=0$.
The sum over the $x_{k}$ is weighted by $C_{k}=\left(\sum^{K}_{j=1;j\ne k}sim\left(e_{j},e_{k}\right)x^{*}_{j}\right)=\left(\sum^{K}_{j=1;j\ne k}sim\left(e_{j},e_{k}\right)\left[2x_{j}-1\right]\right).$

Similarly, for $\mu$, the following is obtained 
\[
\frac{\partial}{\partial\mu}\ln\left(1+\exp\left[-\beta C_{k}-\mu\right]\right)=\frac{\exp\left[-\beta C_{k}-\mu\right]}{1+\exp\left[-\beta C_{k}-\mu\right]}=\frac{1}{1+\exp\left[\beta C_{k}+\mu\right]}
\]
and 
\[
\frac{\partial}{\partial\mu}\ln\left(1+\exp\left[\beta C_{k}+\mu\right]\right)=\frac{\exp\left[\beta C_{k}+\mu\right]}{1+\exp\left[\beta C_{k}+\mu\right]}
\]

and hence 
\[
\frac{\partial L_{p}}{\partial\mu}=\sum^{K}_{k=1}\left[x_{k}\frac{1}{1+\exp\left[\beta C_{k}+\mu\right]}-\left(1-x_{k}\right)\frac{\exp\left[\beta C_{k}+\mu\right]}{1+\exp\left[\beta C_{k}+\mu\right]}\right]
\]
or 
\[
\frac{\partial L_{p}}{\partial\mu}=\sum^{K}_{k=1}\left[x_{k}P\left(x_{k}=0|x_{j}:j\ne k\right)-\left(1-x_{k}\right)P\left(x_{k}=1|x_{j}:j\ne k\right)\right].
\]

The intuition is the same: $P\left(x_{k}=0|x_{j}:j\ne k\right)$ should
vanish whenever $x_{k}=1$, and $P\left(x_{k}=1|x_{j}:j\ne k\right)$
should vanish whenever $x_{k}=0$, for this gradient to approach zero.

For the second derivatives, let 
\[
f\left(x,a,b\right)=\frac{\exp\left(ax+b\right)}{1+\exp\left(ax+b\right)}=1-\left(1+\exp\left(ax+b\right)\right)^{-1}
\]

and 
\[
\frac{\partial f}{\partial a}=\left(-1\right)\left(-1\right)x\frac{\exp\left(ax+b\right)}{\left[1+\exp\left(ax+b\right)\right]^{2}}.
\]

For the second derivatives of the pseudo-likelihood, we have 
\[
\frac{\partial L_{p}}{\partial\beta^{2}}=-\sum^{K}_{k=1}C^{2}_{k}\frac{\exp\left[\beta C_{k}+\mu\right]}{\left(1+\exp\left[\beta C_{k}+\mu\right]\right)^{2}}
\]
and 
\[
\frac{\partial L_{p}}{\partial\mu^{2}}=-\sum^{K}_{k=1}\frac{\exp\left[\beta C_{k}+\mu\right]}{\left(1+\exp\left[\beta C_{k}+\mu\right]\right)^{2}}.
\]

\section{Example and Results}

To illustrate the integrated Ising-LiRA framework, we applied the
model to a simulated dataset of 500 responses to a constructed-response
item. The responses were generated to mimic fourth-grade students'
answers, with 300 scored as correct and 200 as incorrect. The question
was ``What is the relationship between foxes and rabbits in nature?''
Typical answers were:
\begin{itemize}
\item ``The fox eats the rabbit for food.'' (correct)
\item ``Rabbits are prey animals that foxes hunt.'' (correct)
\item ``In the wild the fox hunts rabbits to get food.'' (correct)
\item ``Foxes and rabbits share a water source.'' (incorrect)
\item ``They both need the same kind of shelter.'' (incorrect)
\end{itemize}
Sentence embeddings were obtained using the multilingual MPNet model 

(\texttt{paraphrase-multilingual-mpnet-base-v2}), which produced 768-dimensional
vectors for each response. The average encoding time was approximately
3 seconds on a standard hardware laptop.

A Python implementation of the full estimation procedure, including
embedding generation, similarity computation, and parameter estimation
via maximum pseudo-likelihood, is available from the author. The code
uses the limited memory Broyden--Fletcher--Goldfarb--Shanno algorithm
(L-BFGS; e.g. Liu \& Norcedal, 1989) for estimation and has been tested
with cases up to 6000 responses. This algorithm should work well with
even larger samples, as the logistic regression reduces to a 2 parameter
problem. The only memory intense step is the calculation the $K\left(K-1\right)/2$
similarities $sim\left(e_{j},e_{l}\right)$ for $1\le j<l\le K$.

\subsection{Similarity and Feature Distributions}

The pairwise cosine similarity matrix of the embeddings exhibited
a minimum value of $-0.033$ and a maximum off-diagonal value of $1.000$,
indicating that most responses had non-negative similarity, consistent
with the assumption that embedding spaces rarely produce highly negative
similarities for text in the same language. The pseudo-likelihood
feature $C_{k}=\sum_{j\neq k}\text{sim}(e_{j},e_{k})\,x^{*}_{j}$
had a mean of $96.41$, a standard deviation of $30.87$, and ranged
from $-28.03$ to $127.26$. The predominantly positive values reflect
the tendency of responses to have many similar neighbors with the
same score, which is expected when the scoring process is consistent.

\subsection{Model Estimation}

The parameters $\beta$ and $\mu$ were estimated via maximum pseudo-likelihood
using logistic regression, as described in Section 2.3. The estimates
were 
\[
\hat{\beta}=0.198,\qquad\hat{\mu}=-19.781.
\]
The positive value of $\beta$ confirms that higher linguistic similarity
between two responses increases the probability that they receive
the same score. The large negative intercept $\mu$ reflects two things
\begin{enumerate}
\item the expected value of the similarity weighted sum of scored responses
\item the overall distribution of correct and incorrect scores
\end{enumerate}
While we have 300 correct and 200 incorrect responses, the distribution
of similarities is such that the average similarity between correct
response is 0.76, the average similarity between incorrect reponses
is 0.62, and the average similarity between incorrect and correct
responses is still 0.56. Therefore, the expected value of the sum
of similarites is positive, and the negative intercept $\mu$ shifts
the log-odds toward 0. This is consistent with the model's ability
to capture and control for the interplay between semantic similarity
and the baseline tendency of the scoring process, as well as the average
of the similarity measures overall.

\subsection{Prediction Performance}

Using the fitted model, we computed the predicted probability of a
correct score for each response and assigned a predicted binary score
by thresholding at 0.5. The predictions were compared with the actual
scores. Table~\ref{tab:metrics} summarizes the agreement metrics.

\begin{table}[h]
\centering \caption{\foreignlanguage{english}{Agreement between actual and predicted scores (multilingual MPNet
embeddings).}}
\label{tab:metrics} %
\begin{tabular}{lc}
\hline 
Metric & Value\tabularnewline
\hline 
Accuracy & 0.922\tabularnewline
Precision & 0.912\tabularnewline
Recall & 0.963\tabularnewline
F1 score & 0.937\tabularnewline
Cohen's $\kappa$ & 0.835\tabularnewline
\hline 
\end{tabular}
\end{table}

The high accuracy (92.2\%) and F1 score (0.937) indicate that the
Ising model, with weights derived solely from embedding similarities
and only two free parameters, can reproduce the original scoring decisions
with high fidelity. The Cohen's $\kappa$ of 0.835 represents substantial
agreement beyond chance, further supporting the model's utility as
a diagnostic tool for scoring reliability. The confusion matrix (not
shown) revealed that the model misclassified only 39 out of 500 responses,
with the majority of errors being false negatives (actual correct
predicted as incorrect), as reflected by the higher recall (0.963)
relative to precision (0.912).

\subsection{Interpretation}

The strong predictive performance demonstrates that the modified Ising
model, when combined with semantic embeddings, effectively captures
the dependencies among scores that arise from the linguistic content
of responses. This provides a model-based justification for the LiRA
reference scoring procedure: the weighted majority vote used in LiRA
can be seen as an approximation to the conditional logistic distribution
derived from the Ising model. Moreover, the estimated parameters offer
interpretable diagnostics: a large $\beta$ suggests that similarity
strongly influences scoring consistency, while the magnitude and sign
of $\mu$ reflect the overall difficulty of the item. These results
support the use of the integrated framework for automated reliability
auditing in large-scale assessments without requiring extensive double-scoring.

\section{Conclusions}

This paper introduced a model-based foundation for the Linguistic-Integrated
Reliability Audit (LiRA) by integrating network psychometrics using
a modified Ising model. In this framework, the pairwise interactions
among binary score variables are directly determined by the cosine
similarities of sentence embeddings derived from the written responses,
and the model contains only two free parameters: the temperature $\beta$,
which scales the influence of semantic similarity, and a bias term
$\mu$, which captures the overall tendency of the scoring process.
The conditional distributions of this model reduce to logistic functions,
showing that LiRA's weighted majority voting over semantic neighborhoods
can be understood as an approximation to the optimal conditional classifier
under the Ising model.

An empirical illustration using 500 simulated responses demonstrated
the practical utility of the approach. With multilingual MPNet embeddings,
the model achieved an accuracy of 92.2\%, an F1 score of 0.937, and
a Cohen's $\kappa$ of 0.835, indicating strong agreement between
the predicted and actual scores. These results suggest that the Ising--LiRA
framework can effectively capture the dependencies among scores that
arise from the linguistic content of responses, even with a minimal
number of parameters. The estimated $\beta$ and $\mu$ provide interpretable
diagnostics: a large $\beta$ indicates that similarity strongly drives
scoring consistency, while $\mu$ reflects item difficulty, average
embedding based similarity (average weight) and the baseline probability
of a correct score.

Several advantages of the proposed approach are noteworthy. First,
it uses the full set of responses without requiring extensive double-scoring,
thus reducing operational costs in large-scale assessments. Second,
the probabilistic nature of the model supports uncertainty quantification,
missing-label imputation via Gibbs sampling, and the generation of
benchmark scores for evaluating rater performance. Third, the integration
with LiRA's scalable methodology offers a unified tool for reliability
assessment that can be applied to international programs such as PIRLS,
PISA, and TIMSS.

Nevertheless, some limitations should be acknowledged. The model assumes
that similarity weights are non-negative, which may not always hold
for certain embedding models or cross-linguistic comparisons. The
quality of the results depends heavily on the chosen sentence embedding;
different models may yield varying degrees of semantic fidelity. The
present illustration used simulated data, and validation with real
assessment responses, including multiple human and automated raters,
is necessary to establish the method's robustness in operational settings.
Finally, the pseudo-likelihood estimation, while computationally efficient,
is an approximation to the full maximum likelihood and may be subject
to bias in small samples.

Future work should extend this framework in several directions. First,
the model could be applied to real constructed-response items from
international assessments to compare its reliability estimates with
those obtained from traditional double-scoring. Second, more sophisticated
activation functions and parameterizations, such as those allowing
for differential weighting of agreements and disagreements, could
be explored. Third, uncertainty measures derived from the fitted model,
such as posterior predictive checks or Gibbs-based imputation, could
be used to identify responses with high scoring ambiguity. Finally,
the approach could be extended to polytomous scores and to multi-language
responses using cross-lingual embeddings, thereby broadening its applicability
in multilingual assessment contexts. In sum, the integration of network
psychometrics and LiRA offers a promising, principled, and practical
avenue for enhancing scoring reliability assessment in educational
measurement.

\section*{References}

Bezirhan, U., Jung, J. Y., \& von Davier, M. (2026). Multilingual
sentence embeddings for linguistic-integrated reliability audit. DOI:10.48550/arXiv.2607.17466.

Cohen, J. (1960). A coefficient of agreement for nominal scales. \emph{Educational
and Psychological Measurement, 20} (1), 37--46.

Epskamp, S., Maris, G., Waldorp, L. J., \& Borsboom, D. (2018). Network
psychometrics. In P. Irwing, T. Booth, \& D. J. Hughes (Eds.), \emph{The
Wiley handbook of psychometric testing: A multidisciplinary reference
on survey, scale and test development} (pp. 953--986). Wiley Blackwell.
DOI:10.1002/9781118489772.ch30

Feng, F., Yang, Y., Cer, D., Arivazhagan, N., \& Wang, W. (2022).
Language-agnostic BERT sentence embedding. \emph{Proceedings of the
60th Annual Meeting of the Association for Computational Linguistics}.
DOI:10.48550/arXiv.2007.01852

Gomaa, W. H., \& Fahmy, A. A. (2013). A survey of text similarity
approaches. \emph{International Journal of Computer Applications,
68}(13), 13--18. DOI:10.5120/11638-7118

Gutmann, M. U., \& Hyvärinen, A. (2012). Noise-contrastive estimation
of unnormalized statistical models, with applications to natural image
statistics. \emph{Journal of Machine Learning Research, 13}, 307--361.

Ising, E. (1925). Beitrag zur Theorie des Ferromagnetismus. \emph{Zeitschrift
für Physik, 31} (1), 253--258.

Jung, J. Y., Bezirhan, U., \& von Davier, M. (2026). Reconceptualizing
Scoring Reliability Through Linguistic Similarity. \emph{Educational
and Psychological Measurement}, 86(4), 738--768. DOI:10.1177/00131644251397428

Liu, D. C.; Nocedal, J. (1989). \textquotedbl On the Limited Memory
Method for Large Scale Optimization\textquotedbl . Mathematical Programming
B. 45 (3): 503--528. DOI:10.1007/BF01589116. S2CID 5681609

Molenaar, P. C. M. (2004). A Manifesto on Psychology as Idiographic
Science: Bringing the Person Back Into Scientific Psychology, This
Time Forever. \emph{Measurement: Interdisciplinary Research and Perspectives,
2}(4), 201--218. DOI:10.1207/s15366359mea0204\_1

Rubin, D. B. (1987). \emph{Multiple Imputation for Nonresponse in
Surveys}. Wiley.

von Davier, M. (2018). Diagnosing Diagnostic Models: From von Neumann’s
Elephant to Model Equivalencies and Network Psychometrics. \emph{Measurement:
Interdisciplinary Research and Perspectives,} 16(1), 59--70. DOI:10.1080/15366367.2018.1436827

Wang, L., Yang, N., Huang, X., Jiao, B., Yang, L., Jiang, D., Majumder,
R., \& Wei, F. (2024). Multilingual E5 text embeddings: A technical
report. \emph{arXiv preprint arXiv:2402.05672}. 
\end{document}